\def\simless{\mathbin{\lower 3pt\hbox
             {$\rlap{\raise 5pt\hbox{$\char'074$}}\mathchar"7218$}}}    %~<

\def\simmore{\mathbin{\lower 3pt\hbox
             {$\rlap{\raise 5pt\hbox{$\char'076$}}\mathchar"7218$}}}    %~>

\documentclass[print, structabstract]{aa}

\usepackage{graphicx}
\begin{document}
\title{The spectral and beaming characteristics of the anomalous X-ray
pulsar 4U 0142+61}

\subtitle{}

\author{J. E. Tr\"{u}mper\inst{1}, N. D. Kylafis\inst{2,3}, 
\"{U}. Ertan \inst{4},
\and
A. Zezas\inst{2,3}}

\institute{Max-Planck-Institut f\"{u}r extraterrestrische Physik, 
Postfach 1312, 85741 Garching, Germany\\
\and
University of Crete, Physics Department \& Institute of Theoretical \& 
Computational Physics, 71003 Heraklion, Crete, Greece\\
\and
Foundation for Research and Technology-Hellas, 71110 Heraklion, Crete, Greece\\
\and
Faculty of Engineering and Natural Sciences, Sabanc\i\ University, 
34956, Orhanl\i, Tuzla, \.Istanbul, Turkey
}

\date {Received ; accepted }

%\abstract{}{}{}{}{} 
% 5 {} token are mandatory

\abstract 
% context heading (optional)
{Anomalous X-ray pulsars (AXPs) and soft gamma-ray repeaters (SGRs) constitute
a special population of young neutron stars, which are thought to be
magnetars, i.e., neutron stars with super-strong magnetic fields 
$(10^{14} - 10^{15})$ G.
}
% aims heading (mandatory)
{Assuming that AXPs and SGRs accrete matter from a fallback disk, we attempt
to explain the energy-dependent pulse profiles exhibited by the AXP 4U 
0142+61, as well as its phase-dependent energy spectra.  
}
%% methods heading (mandatory)
{We test the hypothesis that not only the X-ray spectra, but also the 
energy-dependent pulse profiles of 4U 0142+61 are produced by accretion along
dipole magnetic field lines of strength $10^{12} - 10^{13}$ G at the 
neutron-star surface.
}
%% results heading (mandatory)
{In the fallback disk model, the Thomson optical depth along the 
accretion funnel is significant and bulk-motion Comptonization operates
efficiently.  This is enhanced by resonant cyclotron scattering.
The thus scattered photons escape mainly sideways and  
produce a fan beam, which is detected as a main pulse up to energies
of $\sim 160$ keV. 
The approximately isotropic emission from the stellar 
surface (soft thermal photons and reflected hard X-ray ones) is detected
as a secondary pulse.
This secondary pulse shows a bump at an energy of about
60 keV, which may be interpreted as resonant cyclotron scattering 
of fan-beam photons at the 
neutron-star surface.  This implies a dipole
magnetic field strength $B \approx 7 \times 10^{12} (1+z)$
G, where $z$ is the gravitational redshift.
}
%% conclusions heading (optional) 
{Our model explains not only the soft and hard X-ray spectra of the
AXP 4U 0142+61, but also its energy dependent pulse profiles.  If our 
interpretation of the energy dependence of the secondary pulse is correct,
the surface dipole magnetic field strength is comparable to that of X-ray
pulsars.  Much like our Sun, the surface multipole magnetic field strength
may be two orders of magnitude larger, thus allowing for energetic bursts
to occur.  The accretion process is mediated by the dipole field and is not 
interfering at all with the multipole field.
}


\authorrunning{Tr\"{u}mper et al. 2010}
\titlerunning{Dipole field strength of AXP 4U 0142+61} 

\maketitle

%________________________________________________________________

\section{Introduction}

Anomalous X-ray pulsars (AXPs) and Soft Gamma Ray Repeaters (SGRs) are
young neutron stars with X-ray luminosities much larger than their
spin-down power and long periods in the range $2-12$ s. They are widely
believed to be magnetars deriving their X-ray emission from the decay
of super-strong magnetic fields 
($ \simmore ~ 10^{15}$ G) (e.g. Duncan \& Thompson 1992;
Thompson \& Duncan 1995). At the outset, the magnetar model was
developed to explain the giant bursts and the large bursts of SGRs,
which exceed the Eddington limit of neutron-star luminosities by a very large
factor. If the observed spin-down of these sources is interpreted as
the consequence of magnetic dipole braking, the resulting polar field
strengths are of the order of $\simmore ~ 10^{15}$ G,
in qualitative agreement with
what has been inferred from the observed luminosities
(Kouveliotou et al. 1999).  Later on, it was
discovered that AXPs show short bursts as well, though less frequent
and less energetic, leading to the general notion that both types of
sources are closely related or represent even a single class. 
In the magnetar picture, the
steady X-ray emission of these sources, which have luminosities of
typically a few times $10^{35}$ erg s$^{-1}$, 
is thought to be caused by a twist of the
magnetosphere leading to the amplification of the magnetic field and
the acceleration of particles, which produce the X-ray emission. The
twist is caused by rotational motions of a crust plate and has a
lifetime of $\simmore ~ 1$ yr (Beloborodov \& Thomson 2007). 
For recent reviews of
the magnetar model see Woods \& Thompson (2006) and Mereghetti (2008).

An alternative energy source for the persistent and transient X-ray
luminosities of AXPs and SGRs is accretion from fallback disks, first
proposed, but soon abandoned, by van Paradijs et al. (1995), followed by
Chatterjee et al. (2000) and Alpar (2001). This class of models was
developed further in a series of papers  
(Ek\c{s}i \& Alpar 2003; Ertan \& Alpar 2003; Ertan \& Cheng 2004; 
Ertan et al. 2006; Ertan \& \c{C}al{\i}\c{s}kan 2006; Ertan et al. 2007; 
Ertan \& Erkut 2008, Ertan et al. 2009), which make the fallback-disk 
idea quite attractive.
The fallback-disk model gets support from the discovery
of IR/optical radiation from two of the AXPs, 4U 0142+61 (Wang et
al. 2006) and 1E 2259+586 (Kaplan et al. 2009), which 
has been successfully interpreted as
disk emission.  Further indirect support comes from the discovery of three 
planets orbiting a neutron star (Wolszczan \& Frail 1992; Wolszczan 1995).
The fallback-disk model explains the spin-down 
of AXPs and SGRs by the
disk-magnetosphere interaction and requires only ``normal" neutron-star
dipole fields ($10^{12} - 10^{13}$ G). 
In addition, it is successful in predicting the period
clustering of AXPs/SGRs in the range of $2 - 12$ s.  
On the other hand, this model cannot
explain the super-Eddington bursts, which are relatively rare. They are
attributed to magnetar-type activities occurring in local {\it multipole}
fields (star-spots).  

The main physical questions addressed in the
present paper are: 

1) What is the source of energy powering the
persistent emission of these sources? Magnetic-field decay or
accretion?  

2) How large are the polar dipole-field strengths of
these neutron stars?  Are they super-strong ($\simmore ~ 10^{15}$ G ) or ``normal"
($10^{12} - 10^{13}$ G)?  

The spectra of AXPs and SGRs in the range $0.5 - 10$ keV
have been modeled by a superposition of blackbody and power-law
or, alternatively, double blackbody emission. Recently, hard
X-ray tails at energies reaching up to more than 100 keV have been
observed.  For a recent review see Mereghetti (2008).
The luminosities in these tails observed at energies $\simmore ~ 20$
keV are of the same order as the soft X-ray luminosities, which is a
challenge to any model.  Very recently, Enoto et al. (2010) published a 
homogeneous set of data on 7 sources showing both soft and hard components, 
as well as their periods and period derivatives.

Using the bulk-motion and thermal
Comptonization (BMC/TC) model of Farinelli et al. (2008), Tr\"umper et
al. (2010) have shown that accretion from a fallback disk
can produce the hard X-ray ($\simmore ~ 2$ keV) spectrum of 4U 0142+61.  The
mechanism operates close to the hot thermal mound near the bottom of the
accretion column, where the free-falling matter is stopped.  The
required seed photons are produced in two places (see Fig. 1): 
a) The photosphere of the neutron star, which emits most of the low-energy
photons in the $0.5 - 2$ keV range
and  b) the hot thermal mound at the bottom of the accretion
column (e.g., Basko \& Sunyaev 1976; Becker \& Wolff 2007), 
which has a higher temperature than the photosphere.  
In the following, we discuss
the formation of the different spectral components of 4U 0142+61 and
show that not only the observed broad-band spectrum, but also the
energy-dependent pulse profiles (den Hartog et al. 2008) can be
explained by the BMC/TC model. In particular, we present 
evidence that the magnetic-dipole field at the surface of this neutron
star is $\sim 8 \times 10^{12}$ G, consistent with the expectations of the
fallback-disk model.

\section{The model}

In what follows, we give the basic picture that we have in mind, part of which
was presented in Tr\"umper et al. (2010).  In particular, we discuss the 
geometrical constraints that the observations of 4U 0142+61
place on the emission region and the accretion column and then
we discuss the role that the dipole magnetic field plays in the formation
of the X-ray spectrum and the energy-dependent pulse profiles.

\begin{figure}
\centering
\includegraphics[angle=0,width=9cm]{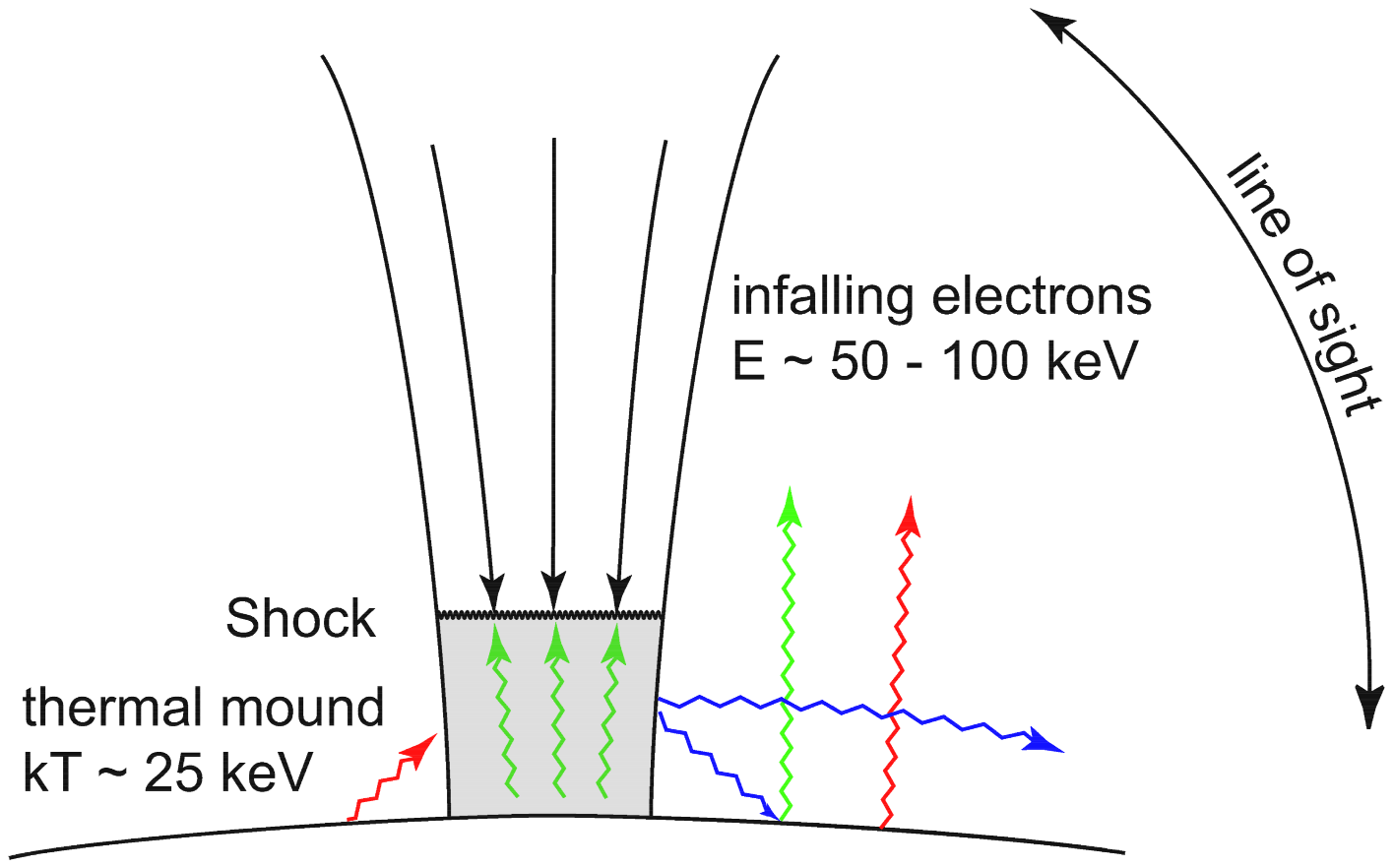}
\caption{
Schematic description of the beaming pattern. The different
wiggly arrows show the mean direction of the respective component.
Blue: The fan beam (main pulse), consisting of photons with energies
between $\sim 2$ and 160 keV, is produced by the BMC/TC process.
It is subject to gravitational bending.  Green:  Fan-beam photons 
hitting the neutron-star photosphere will be reflected (scattered).
Red:  Soft photons from the photosphere escape, while a small fraction 
contributes as seed photons for the Comptonization process.
}
\label{Fig1}
\end{figure}

\subsection{Geometrical constraints for the emission region and optical depths}

Fig. 1 shows a schematic picture of the X-ray emitting region. 
Following Tr\"umper et al. (2010), we assume that the hard X-ray 
component ($2 - 160$ keV) is produced by
thermal Comptonization (TC, $2 - 10$ keV) in the 
accretion shock and the hot mound at the 
bottom of the accretion funnel
and bulk-motion Comptonization (BMC, $\simmore ~ 10$ keV) 
in the accretion flow.

Despite the fact that the magnetic field plays a significant role in the 
radiative transfer 
(see Section 2.2 below), it is instructive to look at the Thomson
optical depths involved in this problem.
In the free-fall regime, the transverse Thomson optical depth $\tau_t$ in the 
direction perpendicular to the magnetic field, near the neutron-star surface,
is
$$ 
\tau_t= n_0 \sigma_T a_0 = 
{{\dot M} \over {m_p v_{ff} \pi a_0^2}} \sigma_T a_0 ,
\eqno(1)
$$
where $a_0$ is the radius of the accretion funnel at the neutron-star radius,
$n_0$ and $v_{ff}$ are the electron density and the free-fall velocity
at the bottom of the accretion funnel, $\dot M$ is the accretion rate,
$\sigma_T$ is the Thomson cross section, and $m_p$ is the proton mass. 
For $\dot M \approx 2 \times 10^{15}$ g s$^{-1}$ 
(corresponding to an X-ray luminosity of $3.7 \times 10^{35}$ erg s$^{-1}$) and 
$v_{ff} \approx 0.5 ~ c$, we get for the transverse optical depth 
$\tau_t \approx 1.7 \times 10^4 / a_0$,
where $a_0$ is measured in cm.  
In order to have the BMC process work efficiently, the BMC region
has to be moderately optically thick in the transverse direction, which 
implies that $a_0$ must be of the order of 100 m or less.
On the other hand, 
the blackbody radius $r_{bb}$ of the observed soft component
$(0.5 - 2)$ keV emitted by the photosphere is much larger. 
It was found to be 8.6 km (White
et al. 1996), $5.4 - 8.3$ km (Israel et al. 1999) or 7.2 km (Juett et
al. 2002). All these radii are normalized to a distance of 3.6 kpc.

In the direction parallel to the magnetic
field the Thomson optical depth of the free-fall zone (the BMC zone) is
$$
\tau_p = {2 \over 3} \tau_t {R \over a_0} \approx 1.1 \times 10^{10} / a_0^2.
\eqno(2)
$$
For $a_0 \simless 100$ m, we obtain $\tau_p \simmore 100$.
In this direction the Thomson optical depth is huge
and the accreting matter acts like an impenetrable wall to the outgoing
photons.  Because of this, we 
can find an upper limit to the transverse Thomson optical depth $\tau_t$ as 
follows:

The hard X-ray spectrum, say from 10 to 100 keV, is produced by repeated 
``head-on" collisions (Mastichiadis \& Kylafis 1992)
between the photons that come out of the 
thermal mound and the infalling
electrons, whose Lorentz $\gamma$ is approximately 1.15
near the neutron-star surface.  The number of such 
collisions required to upscatter an $E_i \sim 10$ keV photon to an
$E_f \sim 100$ keV one is $N_{sc} \sim \log (E_f/E_i) / \log \gamma^2 
\sim 8$.  On the other hand, the observed photon-number spectral index 
of the hard X-rays is $\Gamma \approx 1$.  This means that after each 
head-on collision, roughly 25\% of the photons escape, while the rest suffer
yet one more head-on collision with the infalling electrons.  

Let now $H$ be the mean free path of the outgoing thermally Comptonized
photons in the infalling plasma close to the neutron-star surface.  Then,
$H \approx 1/(n_0 \sigma_T) = a_0/\tau_t$.  The upscattered photons escape 
sideways, through a cylindrical surface of height $\sim H$.  The 
fractional solid angle that this surface subtends at the axis of the 
accretion funnel is $\Omega_{\rm esc} /2 \pi \sim H / a_0 \sim 1 / \tau_t$.
Equating this with the escape probability of $\sim 0.25$, we find 
$\tau_t \sim 4$.

\subsection{The role of the dipole magnetic field}
   
The Farinelli et al. (2008) BMC/TC mechanism applied by Tr\"umper et
al. (2010) has been designed for the non-magnetic case, since it is using
Thomson cross sections. Since we assume that the polar magnetic fields
of the AXPs and SGRs are of the order of $10^{12} - 10^{13}$ G, magnetic cross
sections (e.g. Meszaros 1992) should be used.  After the pioneering
work of Basko \& Sunyaev (1976), 
many people have worked in this area.  
Becker \& Wolff (2005, 2007) have advanced
this complicated undertaking by an analytical approach using angle
averaged magnetic cross sections and a cylindrical approximation for
the accretion column. However, their results are not directly
applicable to AXPs/SGRs, because they have been obtained for sources
like Her X-1 and LMC X-4, which have higher luminosities by more than a
factor of a hundred.  

The most important effect of the strong magnetic field is
that the cross section of the extraordinary-mode photons is significantly
different from the Thomson cross section.  At photon energies $E$ much
smaller than the cyclotron energy $E_c = (h/2\pi) (eB/m_ec)$, the cross
section becomes much smaller than $\sigma_T$ for photons propagating parallel
to the magnetic field, while it becomes very large close to the
cyclotron resonance. These effects are softened somewhat by the
effects of vacuum polarization and spin-flip scattering. We expect
that the inclusion of the magnetic effects will lead to a change of the values
of the parameters $a_0$, $\tau_t$, and $\tau_p$ (Section 2.1).
However, the inclusion of these effects will not change the main result
of Trumper et al. (2010), namely the fact that the broad-band ($0.5 - 200$ 
keV) spectrum of 4U 0142+61 can be represented by a single physical model
involving TC and BMC of photons by the infalling electrons.  Any
deviation from the geometry of the TC/BMC model of Farinelli et al. (2008)
and the inclusion of effects due to a strong magnetic field will be 
absorbed in the values of the model parameters and as a result 
the later should 
not be taken at face value.  We expect that the parameters related to 
the seed photon spectrum (temperature and slope of the modified 
blackbody, and the illumination factor) will be mostly affected by the 
change in the geometry discussed in Section 2.1 (Fig. 1).

Extraordinary photons of energy
$E$ will be scattered quasi-isotropically by electrons when passing
sheets in which the resonance condition 
$E \approx E_c = \gamma (h/2\pi) (eB/m_ec)$ is
fulfilled, where $\gamma$ is the Lorentz factor of the electrons 
($\gamma \approx 1.15$ for free-falling electrons near the neutron-star surface
and $\gamma = 1$ for thermal electrons). 
For photons of energy $E$, the scattering sheet is located at a radius
$r(E) = R (E_{c0}/E)^{1/3}$, where $E_{c0}$ is the cyclotron resonance 
energy at the neutron-star surface.

The resonant scattering optical depth in the accretion column, parallel to
the magnetic field, greatly exceeds the
optical depth for Thomson scattering (Lyutikov \& Gavriil 2006)
$$
\tau_{\rm res} \sim {{ \pi c} \over {8r_e \omega_c}} 
\sim 10^5 ~ \left( { {1 ~ {\rm keV}} \over {E_c} } \right),
\eqno(4)
$$             
where $r_e$ is the classical electron radius, and $\omega_c$, 
$E_c$ are the cyclotron frequency and cyclotron energy,
respectively. This means that, for photon energies in the range $0.2 -
160$ keV, cyclotron resonance scattering is the main source of
opacity. In the ``resonance layer", located at radius 
$r(E) \approx R (E_{c0}/E)^{1/3}$, photons of energy $E$ are 
quasi-isotropically redistributed. 

In addition,
the scattering by the cyclotron harmonics has to be taken into
account, which for a given photon energy $E$ takes place at radii
$r(E) \approx R(jE_{c0}/E)^{1/3}$, with $j = 2, 3, 4 ...$. 

Since the dipole magnetic field decreases rapidly in the vertical direction, 
the cyclotron scattering dominates over a wide range of photon energies
and strongly prevents extraordinary photons from 
traveling upstream.  At the same time, these scattering
effects enhance the efficiency of the bulk-motion Comptonization process.

For extraordinary mode photons, traveling at small angles with respect to the
magnetic field direction, with energies much smaller than the local
cyclotron energy, the scattering cross section becomes much smaller than the
Thomson value.  This appears to have the opposite effect than the one 
discussed above.  However, the extraordinary photons traveling upwards 
will eventually be stopped by cyclotron scattering when passing a 
resonance layer at higher heights.

On the other hand, the ordinary-mode photons also cannot escape in the
magnetic field direction, because the Thomson optical depth 
$\tau_p$ is significantly larger than one and, in addition,
the photons are advected downward by the accretion flow. Since $\tau_p$ 
is significantly larger than $\tau_t$, nearly all the ordinary-mode
photons escape sideways, i.e., perpendicular to the accretion flow.
In view of the above, we conclude that the column will largely emit a fan
beam, as shown schematically in Fig. 1.

\subsection{Interpretation of the 4U 0142+61 pulse shapes and energy spectra}

A comprehensive and detailed description of the broad-band X-ray
characteristics of AXP 4U 0142+61 has been given by den Hartog et
al. (2008).  In the following, we present a detailed discussion of these
results in terms of the accretion model.  To this end, we use two results
of the den Hartog et al. (2008) paper, the energy-dependent pulse profiles
shown in Fig. 2, measured by XMM-Newton, RXTE-PCA, and INTEGRAL-ISGRI, 
and the complementary phase-dependent energy spectra shown in Fig. 3.

\begin{figure}
\centering
\includegraphics[angle=0,width=9cm]{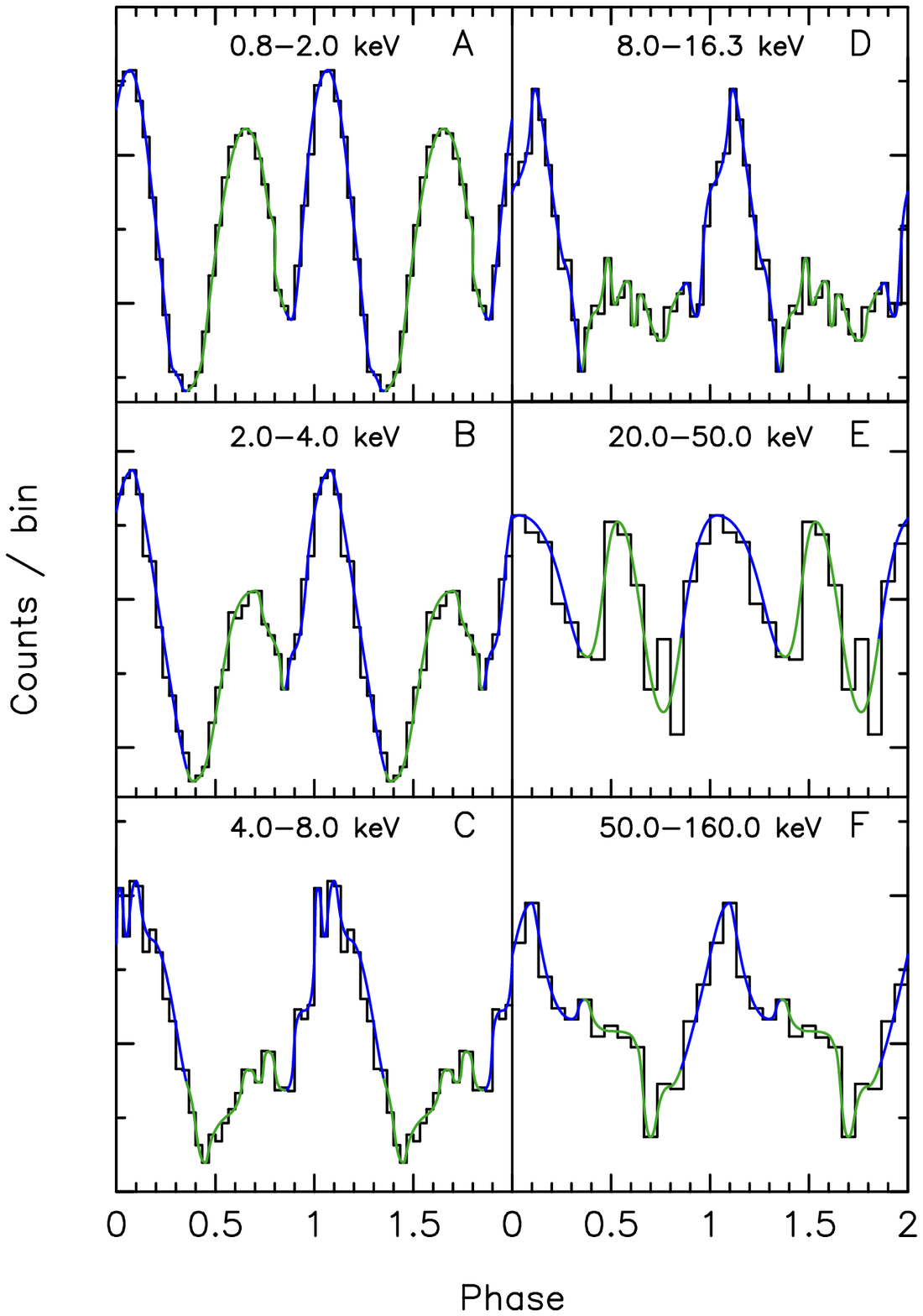}
\caption{
Schematic presentation of the energy dependent pulse profiles
measured by XMM-Newton, RXTE-PCA and INTEGRAL-ISGRI. The main pulse 
(blue phase) ranges from the lowest to the highest energies. In
our model, this component is mainly formed by the fan-beam photons
produced by the BMC/TC process.
The secondary pulse (green phase)
is produced by the photospheric radiation at low energies ($\simless ~ 2$ keV)
and the fan-beam photons, which hit the photosphere and are reflected there.
The intensity of this pulse falls off quickly between 0.8 and 8 keV.  
It reappears at higher energies and becomes as prominent as the main
pulse in the $16.3 - 50$ keV interval.
This figure is a simplification of Fig. 7 of
den Hartog et al. (2008).
}
\label{Fig2}
\end{figure}

\begin{figure}
\centering
\includegraphics[angle=0,width=9cm]{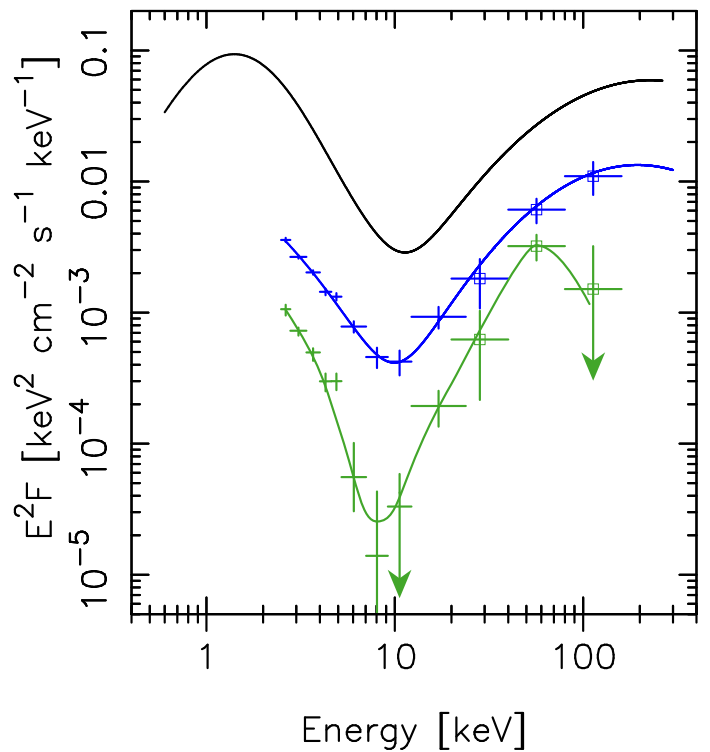}
\caption{
This figure is a simplification of Fig. 8 of den Hartog et al. (2008),
from which the data are taken.
The black line represents the INTEGRAL/XMM-Newton total spectrum fit 
shown in that paper.  The spectrum of the main pulse is shown in blue.
The spectrum of the secondary pulse in shown in green.  
In our model, this component consists above $\sim 8$ keV of fan-beam
photons reflected by the photosphere.  Its spectrum shows a rapid
rise up to energies of $\sim 60$ keV and a break at $\sim 80$ keV.  
In our model, this bump, which can also be seen in the pulse shapes
(Fig. 2), is interpreted in terms of enhanced reflection due to cyclotron
resonance scattering.
}
\label{Fig3}
\end{figure}

As noted by den Hartog et al. (2008), morphology changes are ongoing
throughout the whole energy range.  However, the principal
structure of the pulse profiles is very simple:
It mainly consists of two pulses (c.f. Fig. 2).  
In Figs. 2 and 3 we present simplified versions of the
energy-dependent pulse profiles and the phase-dependent spectra,
which are labeled by colors.
The main pulse (blue color), peaking
around phase 0.1, is located at a stable position from the lowest to
the highest energies and the secondary pulse (green color)
is located at phases around 0.6 (Fig. 2).  
The phase-dependent spectra (Fig. 3) show the same color.

\subsubsection{Interpretation of the main pulse}

The interpretation of the main pulse is very simple
in our model.  We identify it with the fan beam emitted 
from the region of the hot shock/mound 
and the infalling matter above it (Fig. 1).
At the lowest energies $(E ~ \simless ~ 1.7)$ keV, 
the emission consists of two components, the
blackbody-like emission from the part of the extended photosphere that
faces the observer and the low-energy tail of the BMC/TC spectrum that
is emitted as a fan beam.
In the energy interval $1.7 ~ \simless ~ E ~ \simless ~ 10$ keV, 
the spectrum is composed of the tail 
of the photospheric spectrum and the power-law spectrum produced by the 
thermal Comptonization (TC).
At energies $E ~ \simmore ~ 10$ keV, 
up to the highest energies beyond $100$ keV,
the emission is due to the bulk-motion Comptonization (BMC).

\subsubsection{Interpretation of the secondary pulse}

In Figs. 2 and 3 we show as green lines the secondary pulse, 
which occurs at phase $\sim 0.6$.  We identify it with the polar beam (Fig. 1).
As in the case of the main pulse, here also we 
consider three energy bands.

The intensity of the secondary pulse rises strongly towards low energies
(Fig. 2).  An extrapolation of the trend shows that
at energies below 0.8 keV it becomes stronger than the main pulse, which
is fully consistent with the ASCA data in the
$0.5 - 1.7$ keV band (Kuiper et al. 2006; den Hartog et al. 2008).  
We interpret this component 
in terms of the blackbody-like spectrum dominating at energies below 2 keV
and originating in the extended hot photosphere
surrounding the accretion column, which has a rather large extent 
($r_{bb} \sim 5 - 8$ km).  
The main difference in the angular distribution compared with that of a 
blackbody is that a broad peak occurs in the polar direction, caused by
extraordinary photons escaping from larger depths (c.f. Pavlov et al. 1994).
The emission from this area will be composed of an approximately isotropic
component and a narrower pencil beam of extraordinary-mode photons,
which escape from the deeper layers along the magnetic field lines.

In the $1.7 - 10$ keV band, the intensity of the secondary pulse falls off
steeply with energy.  In our model, this part of the spectrum is produced 
by the superposition of the tail of the blackbody distribution and the
power-law spectrum produced by thermal Comptonization (TC).  
A possible third component can be produced by reflection of a fraction of
fan-beamed TC photons, which hits the polar-cap area.
The reflected photons leave the neutron-star surface more or less 
isotropically, i.e. with an angular distribution similar to that of the
thermal emission from the photosphere.  Thus, both components are in phase.

In the $10- 50$ keV band, the secondary pulse becomes quite strong and it is
actually comparable in strength with the main pulse.  We note that this pulse
extends into the $50 - 160$ keV energy band, where it is visible as a 
shoulder of the main pulse (Fig. 2).  But the effect is small
due to the large width of this energy interval.  
It can be seen better if
one looks at the phase-resolved spectra shown in Fig. 3.
The secondary pulse
spectrum shows a broad bump in the $20 -80$ keV range.  Actually, its 
intensity rises faster than that of the main pulse up to an energy of 
$\sim 60$ keV, beyond which it indicates a drop.
We will return to this feature in Section 3.
In our model, this part of the spectrum is produced by the BMC photons that
hit the neutron-star surface and get reflected (scattered)
with some energy loss.  Since the reflected photons have a more or less
isotropic distribution in direction, this energy band is in phase with
the previous two.

\subsubsection{Remarks}

Concluding our interpretation of the energy-dependent pulse shapes, 
we think it is appropriate to give some remarks:

1.  In Fig. 1, we show the swing of the line of sight during half a 
phase cycle, approximately required by our model for 4U 0142+61,
which ranges from 
a direction roughly perpendicular to the accretion column (main pulse)
to an off-pole direction (secondary pulse).  We stress that 
the phenomenological appearance of an AXP or SGR in terms of 
energy-dependent pulse shapes or phase-dependent spectra will depend 
crucially on this viewing geometry, which in turn depends on the angles 
between the rotational axis of the neutron star and the line of sight 
on one hand and the dipole axis, on the other.

2.  The gravitational bending of the fan beam may play an important role
for the beaming, depending on the unknown mass and radius of the
neutron star.

3.  As mentioned above, the hot polar cap has a radius $r_{bb} \sim 5 - 8$ km.
The spectral fits of White et al. (1996), Israel et al. (1999), and Juet
et al. (2002) yield a temperature of $kT \approx 0.4$ keV.  However, this 
is probably a flux-averaged temperature, since the temperature will decrease 
from higher values close to the accretion column to lower values at
larger distances.

4.  The energy-dependent pulse shapes in Fig. 2 represent only the pulsed
components without the constant components.  The pulsed fractions can be
found in Fig. 9 of den Hartog et al. (2008).  They are quite low at soft
energies ($\sim 10\%$ at 1 keV), which is qualitatively expected from our 
model, since the photosphere is visible to the observer all the time 
(see Fig. 1).  On the other hand, the TC/BMC components show larger pulse 
fractions ($\sim 25\%$ above 28 keV), as anticipated for the dominating 
fan beam.

\section{An estimate of the magnetic dipole field of AXP 4U 0142+61?} 

We believe that the dipole magnetic field is not only channeling the accretion
flow onto the poles, but also leaves its imprint in the observed spectra.  
As already discussed above, the spectrum of the secondary (polar)
beam is characterized by a fast rise (faster than that of the fan beam)
from $\sim 10$ keV up to a maximum at $\sim 60$ keV, 
followed by a decline or a
cutoff beyond $\sim 70$ keV. This behavior is expected if the polar beam is
produced by scattering at the photospheric plasma, whose efficiency rises
for energies approaching the cyclotron energy.
Interpreting the spectrum this way, leads to a cyclotron energy of $70 - 80$
keV, corresponding to a polar magnetic field strength of $B_0 \sim 6 - 7
\times 10^{12} (1+z)$ G, where $z$ is the gravitational redshift. 
Clearly, observations with higher photon statistics would be necessary to
confirm this conclusion.

\section{Heating of the hot polar cap}

The photosphere responsible for the soft component has a flux-averaged 
temperature of $\sim 0.4$ keV and a blackbody radius $r_{bb} \sim 5 - 8$ km. 
In our model, there are three possible ways to heat the polar cap:

1. By absorption of a fraction of the fan-beam photons hitting the neutron-star 
surface.  This process will be enhanced by the gravitational bending of 
the fan beam towards the neutron-star surface, which depends on the gravity
of the neutron star, i.e., its mass over its radius.  If this were the main 
heat source of the polar cap, the power of the re-emitted thermal component 
and the reflected (scattered) component would be equal to the power of the fan 
beam impinging on the polar cap.

2. A second heat source may be that part of the accretional power which is 
conducted into the stellar interior and redistributed over a larger area, 
depending on the magnetic-field configuration in the interior and the crust.

3. Another contribution to the heating of the polar cap could be provided by 
conduction of heat from the hot interior of the relatively young neutron star.

In view of our limited quantitative understanding of all properties and 
processes, which play a role in this context, it would be premature to favor
a specific combination of these processes.  

\section{Summary and conclusions}

We have shown in a semi-quantitative way that bulk-motion/thermal
Comptonization in an accretion column, formed by a dipole magnetic field
of strength $\sim 10^{13}$ G,
describes well not only the soft and hard X-ray spectra but also
the phase-dependent energy spectra of AXP 4U 0142+61. 
The energy-dependent pulse profiles and their constancy over long periods
of time constitute significant observational constraints for the proposed 
models.  Our model explains them in a simple and natural way by the formation
of two ``beams", one perpendicular to the accretion column (main pulse)
and one roughly parallel to it (secondary pulse).
AXP 4U 0142+61 is one of seven sources presently known to have hard X-ray 
tails (Enoto et al. 2010).  From the similarity of their broad-band spectra,
it is no surprise that accretion with thermal and bulk-motion
Comptonization works for all of them (Zezas et al., in preparation).

On the other hand, the magnetar model of a twisted magnetic field 
(Thompson et al. 2000), where the hard X-ray emission originates far 
up in the magnetosphere, seems to have problems in explaining the
energy-dependent pulse profiles (Fernandez \& Thompson 2007)
Another problem for the magnetic-twist model is the long
term stability of the pulse shape and phases. This would require that
the rotation of the
same crustal plate, located at a fixed position on the neutron star,
is responsible for the quiescent emission, which looks quite
artificial. 
Furthermore, the $\dot P$ of SGR 0418+5729 was determined (Rea et al. 2010)
to be less than $6 \times 10^{-15}$ s s$^{-1}$, which in the magnetar model
implies a magnetic dipole field of $< ~ 7.5 \times 10^{12}$ G!

In this paper we propose that the spectral bump of the secondary pulse (which
corresponds to the polar beam) is caused by cyclotron-resonance reflection
of the fan beam by the magnetized photosphere. This leads to a 
magnetic-field strength of $7 \times 10^{12} (1+z)$ G, which
falls in the range of dipole magnetic fields 
($10^{12} - 10^{13}$ G) attributed to AXPs/SGRs in
the framework of the fallback-disk scenario. 

On the other hand, we stress that the few giant bursts 
observed from SGRs 
cannot be explained by any accretion process. The same is
true for bursts with large super-Eddington luminosities. They are most
likely caused by crustal shifts of plates carrying super-strong
magnetic fields ($\simmore 10^{15}$ G), as discussed in the classical magnetar
literature (e.g. Thomson \& Duncan 1995). Our analysis suggests that
these events do not take place in the dipole field, but in 
localized multipole fields. This situation is qualitatively
similar to that of the Sun, which shows flare activities in sunspot fields,
which are larger than the solar dipole field by at least two orders of
magnitude.

In conclusion, our model explains naturally the X-ray spectra and the 
energy-dependent pulse profiles of 4U 0142+61.  Furthermore, it allows
this source or any other AXP or SGR to emit big or giant bursts if the
neutron star involved sustains multipole fields with strengths in the 
$10^{14} - 10^{15}$ G range.

\begin{acknowledgements}
This research has been supported in part by EU Marie Curie project no. 39965, 
EU REGPOT project number 206469 and by EU FPG Marie Curie Transfer of Knowledge 
Project ASTRONS, MKTD-CT-2006-042722. \"{U}.E. acknowledges research support 
from T\"{U}B{\.I}TAK (The Scientific and Technical Research Council of Turkey) 
through grant 110T243 and support from the Sabanc\i\ University Astrophysics 
and Space Forum. 
\end{acknowledgements}

\end{document}